\documentclass[a4paper]{jpconf}

\usepackage{graphicx}

\usepackage{amsmath}

\usepackage{listings}   
\usepackage{url}

\begin{document}

\title{RooStatsCms: a tool for analysis modelling, combination and statistical studies}

\author{D.~Piparo$^1$, G.~Schott$^1$ and G.~Quast$^1$}

\address{$^1$ Faculty of Physics, Karlsruhe University, Wolfgang-Gaede-Str. 1, 76131 Karlsruhe}
\ead{danilo.piparo@cern.ch, gregory.schott@cern.ch, gunter.quast@cern.ch}

\begin{abstract}

RooStatsCms is an object oriented statistical framework based on the
RooFit technology.  Its scope is to allow the modelling, statistical
analysis and combination of multiple search channels for new phenomena
in High Energy Physics. It provides a variety of methods described in
literature implemented as classes, whose design is oriented to the
execution of multiple CPU intensive jobs on batch systems or on the
Grid.

\end{abstract}

\section{Introduction}

Statistical analysis and the combination of measurements has a
dominant importance in High Energy Physics. It is very challenging
from the point of view of the tools to be deployed, the communication
among the analysis groups and the definition of statistical
guidelines. Previous experiments such as those at LEP \cite{LEPEWG} and
Tevatron \cite{TevatronEWG} already devoted huge efforts in this
direction.

At the LHC, early results will require the combined analysis of
different search channels and eventually the combination of results
obtained by different experiments. There will definitely be a need to
build complex models, i.e.~parametrisations, to describe the
experimental distributions, to quantify a possible signal excess in
the data or to set a limit on the signal size in the absence of such
an excess. In addition, a quantitative statistical treatment will
require extensive studies based on Monte-Carlo, and should consider
different statistical methods. The combination of analyses require a
reliable and practical transfer of data and models across working
group and experiment boundaries. Previous attempts to achieve these
goals were built upon dedicated code for each analysis, and a very
tedious and often error-prone transfer of the obtained results into 
the combination procedures.

In order to perform the statistical treatment of combination of
analyses multiple methods are available. The choice of the method to
use depends often on the context of the analysis and on the
interpretation of the data by the experimenter. When multiple methods
are applicable, a comparison of their results might be useful or
is even required. It is therefore important to be able to easily
switch between methods without too much effort. This
was so far not possible as the implementation of different approaches
were not unified in a single package and a comparison would require
the user to learn how to use a number of packages. 

It is the lack of a general, easy-to-use tool that led us to the
decision to develop RooStatsCms (RSC) \cite{RooStatsCms} for analysis
modelling, statistical studies and combination of results. A first
release of the RSC package has been provided in February 2008. It
relies on the ROOT~\cite{ROOT} extension RooFit \cite{RooFit}, from
which it inherits the ability to efficiently describe the analysis
model, thereby separating code and descriptive data, and easily
perform toy Monte-Carlo experiments. A selection of different methods
for statistical analysis and combination procedures is also included.

We start describing the software environment of RooStatsCms in
section~\ref{Framework_and_software_environment}. Since RSC is made up
of three components, i.e.~the modelling part, devoted to the
construction of the analysis model starting from a configuration file,
the implementation of the statistical methods and a set of advanced
graphics routines, a natural way to describe it is to characterise
these parts separately, in sections~\ref{Analyses_modelling},
\ref{Implementation_of_the_statistical_methods} and
\ref{Graphics_routines} respectively. In
section~\ref{Statistical_methods}, we then give an introduction to the
statistical methods before showing some examples of applications of
RSC in section~\ref{Examples_of_applications}.

\section{Framework and software environment}
\label{Framework_and_software_environment}

RooStatsCms is entirely written in \verb|C++| and relies on
ROOT. The ROOT Analysis Framework is the most widely used tool in the
High Energy Physics community for data analysis. It exploits an
advanced object oriented design to reach a high scalability and
flexibility. Among its most important components there are
cutting-edge visualisation tools, a rich set of container classes that
are fully I/O aware, an extensive set of GUI classes, run-time object
inspection capabilities, shared memory support and an automatic HTML
documentation generation facility. The \verb|TObject| class provides
default behaviour and protocol for all objects in the ROOT system. It
provides a protocol for object I/O, error handling, sorting,
inspection, printing and drawing. Every object which inherits from
\verb|TObject| can be stored in the ROOT collection classes or written
to ROOT files on disk. One of the key components of ROOT is the CINT
\verb|C|/\verb|C++| interpreter~\cite{CINT} since it allows rapid
prototyping eliminating the typical time consuming edit/compile/link
cycle.  Existing \verb|C|/\verb|C++| libraries can be easily
interfaced to the interpreter. This is done by generating a dictionary
from the functions and classes definitions which is then compiled and
linked with the code into a single libray. The CINT interpreter is
fully embedded in ROOT allowing command line, scripting and
programming languages to be identical. RSC is distributed with the
CINT dictionaries and its classes and functions can therefore be used
also inside macros or in the interpreter.

To reach a maximum flexibility and exploit all the recent
technologies, we decided to use the RooFit 
toolkit. This package was a project started originally for the
analyses of the BaBar experiment and is now a part of ROOT.
The RooFit technology is based on a cutting-edge object oriented
design, according to which almost every mathematical entity is
represented by a class. For example, parameters and variables are
treated symmetrically and can be expressed as \verb|RooRealVar|
objects, representing real intervals, holding an (asymmetric) error
and a fit range. Also probability density functions (pdf) are
represented by classes inheriting from \verb|RooAbsPdf| and using
their methods very complicated objects can be described. The numerous
simple models provided in the package such as Gaussians, polynomials
or Breit-Wigners can be combined to build the elaborate shapes needed
for the analyses.  Many operations from pdf objects are possible; the
user can perform pdf addition, convolution or product of pdfs of
different variables. The communication among the class instances
bounded together in the complex structures that result from such
operations, is granted by an advanced reference caching mechanism.
The persistency of these composite objects is assured by the
\verb|RooWorkspace| class, which is able to go through these
references and bring to disk all the necessary objects. In any case,
RooFit takes care automatically of the normalisation of the pdfs with
respect to all of their parameters and variables within their defined
ranges.  All the integration procedures are highly optimised,
combining analytic treatment with advanced numerical
techniques. Simultaneous and disjoined fits can be carried out with
the possibility to interface RooFit to \verb|MINUIT| and other
minimisation packages. On the top of that, optimised toy Monte-Carlo
dataset generation is provided.

RooStatsCms is integrated in the official CMS~\cite{cms} Software Framework
\cite{CMSSW}, in the \verb|PhysicsTools/RooStatsCms| package, starting
from the 3.1.X series.

\section{Analyses modelling}
\label{Analyses_modelling}

In the analysis of a physics process
the description of
its signal and background components, together with
correlations and constraints affecting the parameters, is a critical step.  
RooStatsCms
provides the possibility to easily model the analyses and their
combinations through the description of their signal and background
shapes, yields, parameters, systematics and correlations via
analysis configuration files, called \textit{datacards}.  The goal
of the modelling component of RSC is to parse the datacard 
and generate from it a model according to the RooFit
standards. There are a few classes
devoted to this functionality, but the user really needs to deal only
with one of them, the \verb|RscCombinedModel|.
The approach described above has mainly three advantages: 
the factorisation of the analysis description and statistical treatment 
in two well defined steps, 
a common base to describe the outcomes of the studies by the analysis groups, 
and a straightforward and documented sharing of the results.

A datacard is an ASCII file in the ``.ini'' format, therefore
presenting key-value pairs organised in sections. 
This format was preferred to the XML because of its simplicity and high 
readability.

The parsing and processing of the datacard is achieved through an
extension of the RooFit \verb|RooStreamParser| utility class.  This
class is already rather advanced. Beyond reading strings and numeric
parameters from configuration files, it implements the interpretation of
conditional statements, file inclusions and comments.  In presence of
a complicated combination, the user can take advantage in RSC from
these features specifying one single model per datacard file and then
import all of them in a ``combination card''.

Every analysis model can be described as a function of one or many
observables, e.g.~invariant mass, output of a neural network or
topological information regarding the decay products.  For each of
these variables a description of the signal and background case is to
be given, where both signal and background can be divided in multiple
components, e.g.~multiple background sources. To each signal
and background components, a shape and a yield can be assigned. For
what concerns the shape, a list of models is present and for those
shapes which are not easily parametrisable, a \verb|TH1| histogram in
a ROOT file can also be specified.  The yields can be expressed as a
product of an arbitrary number of single factors, for example:
$\mathrm{Yield} = \int{\mathcal{L}} \cdot \sigma \cdot BR \cdot
\epsilon$, where $\int{\mathcal{L}}$ is the integrated luminosity,
$\sigma$ a production cross section, $BR$ a decay branching ratio and
$\epsilon$ is the detection efficiency.

Using RooFit, all the parameters present in the datacard can
be specified as constants or defined in a certain range. In addition
to that, exploiting the RSC implementation of the constraints, the
user can directly specify the parameter affected by a Gaussian or a
log-normal systematic uncertainties. In the former case, correlations can be
specified among the parameters via the input of a correlation matrix.

In a combination some parameters might need to be the same throughout
many analyses, e.g.~the luminosity or a background rate. This feature is
achieved in the modelling through a ``same name, same pointer'' mechanism. 
Indeed every parameter is represented in memory as a \verb|RooRealVar| or, 
in presence of systematic uncertainties, as a derived object, the 
\verb|Constraint| object and the \verb|RscCombinedModel| merges all 
variables with the same name via an association to the same pointer.

\section{Implementation of the statistical methods}
\label{Implementation_of_the_statistical_methods}

\begin{figure}[h]
    \begin{center}
        \includegraphics[width=100mm]{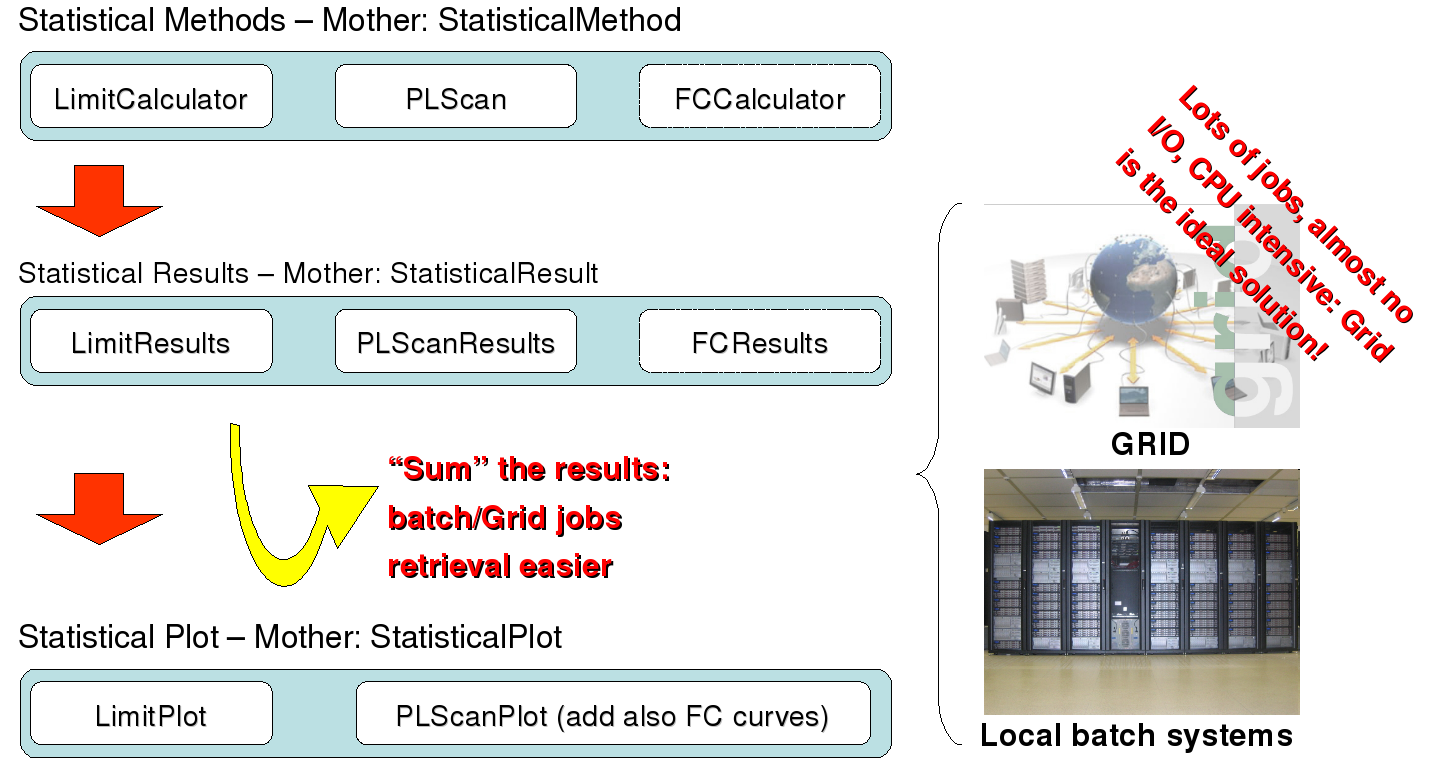}
        \caption{The RSC classes structure, excluding the modelling component. 
                 It is designed to ease job submission and recollection of results.}
        \label{classes_structure}
    \end{center}
\end{figure}

Each statistical method in RooStatsCms is implemented in three classes types 
(this structure is reflected in the code by three abstract classes, 
see figure \ref{classes_structure}): 
the \textit{statistical method} where all the time consuming operations such as 
Monte-Carlo toy experiments or fits are performed,  
the \textit{statistical result} where the results of the computations are 
collected and \textit{the statistical plot} whose role is to provide a 
graphical representation of the \textit{statistical result}. 

In many cases, e.g.~frequentist approaches (see section
\ref{Frequentist_methods}), the CPU time needed for the calculations
can be considerable.  An interesting feature of the
\textit{statistical result} classes is that their objects can be
``summed up'': this is very useful to accumulate statistics when
combining the outputs of many processes.  Indeed, the classes
factorisation described above combined with the persistency of the RSC
objects, eases the submission of jobs to a batch system or to the Grid
and the recollection of the results, allowing to carry out such
calculations at in reasonable timescales.

The \textit{statistical plot} classes play a fundamental role in a 
statistical analysis, providing a graphical representation of the results in 
the form of self explanatory plots. The objects of these classes can 
directly be drawn onto a \verb|TCanvas| via a \verb|draw| method and, when 
needed, all the components used to produce the plot (\verb|TGraph|, 
\verb|TH1F|, \verb|TLegend|, \dots) can be saved separately in a ROOT 
file for a further manipulation.

\section{Graphics routines}
\label{Graphics_routines}

This category of classes is devoted to the production of plots. There
are two types of plots covered: those that summarise information
collected during the running of the statistical classes (such as
figures~\ref{plscan} and \ref{m2lnQ}), and plots allowing a graphical
display of the physics results obtained.
In this second category, if on the one hand, RooStatsCms does not
provide any user-specific graphics routines, however, during the past decade,
the LEP and Tevatron collaborations established a sort of standard to
display the results of (combined) searches for new signals. This kind
of plots are now well accepted in the community, and for this reason
utility routines are provided to produce them. 
Figures~\ref{htautau_lim_1} and \ref{htautau_lim_2} show two examples of 
plots. A few more examples are shown in section~\ref{Examples_of_applications}.

\section{Statistical methods}
\label{Statistical_methods}

We will now present the principles of some of the statistical methods
implemented in the package before illustrating in the next section
results of applications in CMS analyses. For a more detailed
introduction to statistical methods in High Energy Physics see for
example \cite{fredjames}.

\subsection{Profile likelihood approach}
\label{Profile_Likelihood}

Suppose we have, for each of $N$ events in a collection, a set of
measured quantities \mbox{$\underline{x}=(x_a,\,x_b,\,x_c,\,\ldots)$}
whose distributions are described by a joint probability density
function, or pdf, $f(\underline{x},\underline{\theta})$, where
$\underline{\theta}=(\theta_1,\,\theta_2,\,\theta_3,\,\ldots )$ is a
set of $K$ parameters. The likelihood function is then defined by
the equation:
\begin{equation}
    \label{likelihood}
    L(\underline{x},\underline{\theta})=\prod^{N}_{i=1}f({\underline{x}}_{i},\underline{\theta}).
\end{equation}
To ease the calculations, the negative of the logarithm of the likelihood
function $-\ln{L}$ (negative log-likelihood) is often used.

Focussing on a one-dimensional case, the profile likelihood method
requires to perform a scan over a sensible range of values of the
parameter of interest $\theta_{0}$. For every point of this scan, the
value of $\theta_{0}$ is fixed and $-\ln{L(\theta_{0})}$ is minimised
(i.e.~a fit is performed) with respect to the remaining $K-1$
parameters. The maximum likelihood estimator of the $\theta_{0}$
parameter, noted $\hat{\theta}_0$, is the value where the negative
log-likelihood function is at its minimum
$-\ln{L(\hat{\theta}_0)}$. 

\begin{figure}[h]
    \begin{center}
        \includegraphics[width=87mm]{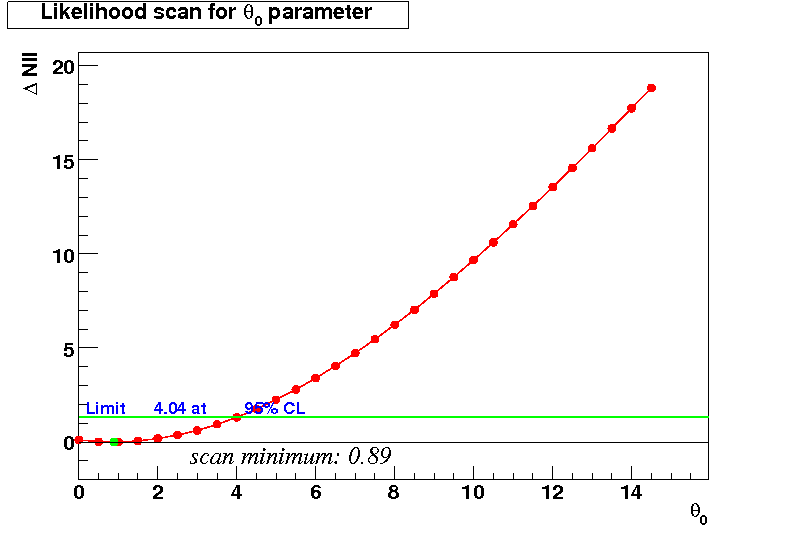}
        \caption{The negative log-likelihood scan over the parameter
	         $\theta_0$. The horizontal line at $\Delta
	         nll\simeq1.36$ allows to read for the 95\% CL upper
	         limit on this parameter.  The interpolation done by
	         RooStatsCms between scan point pairs is linear, while
	         the minimum of the scan is the minimum of a parabola
	         built on the lowest three points.\label{plscan}}
    \end{center}
\end{figure}

Figure~\ref{plscan} shows an example of a profiled negative
log-likelihood curve which has been offset by
$-\ln{L(\hat{\theta}_0)}$. From this construction, it is possible to
easily obtain the one- or two-sided confidence interval we are
interested in. In the assumption of a parabolic shape of the negative
log-likelihood function\footnote{It can be shown that this approach is
also valid for a general scan shape \cite{Metzger}.}, the boundaries
of a confidence intervals correspond to the values of
$\theta_0$ with
\begin{equation}
    \label{parabolicnll}
    \Delta nll = -(\ln{L(\theta_0)}-\ln{L(\hat{\theta}_0)}) = \frac{n_\sigma^{2}}{2},\ \mathrm{with}\ n_\sigma=\frac{\theta_{0}-\hat{\theta}_{0}}{\sigma}.
\end{equation}
where $\sigma$ represent the Gaussian standard deviations. The mapping between the desired 
confidence level (CL) and the value of $n_\sigma$ is given, in the Gaussian assumption, by the formulae:
\begin{eqnarray}
    \label{nsigmasfromCL1}
    &n_\sigma = \sqrt{2}\cdot \mathrm{Erf}^{-1}(CL) & \mathrm{(two-sided)} \\
    &n_\sigma = \sqrt{2}\cdot \mathrm{Erf}^{-1}(2 \cdot CL -1) & \mathrm{(one-sided)}
    \label{nsigmasfromCL2}
\end{eqnarray}
where $\mathrm{Erf}^{-1}$ is the Gaussian inverse error function. In
the later case, an upper limit on the parameter $\theta_0$ would then
correspond then to the value $\theta_0>\hat{\theta}_0$ obeying
equations~\ref{parabolicnll} and~\ref{nsigmasfromCL2}.

\subsection{Frequentist approach}
\label{Frequentist_methods}

Analysis of search results can be formulated in terms of hypothesis
testing in a frequentist approach (for an explanation see
\cite{alexread}). We define $H_b$ as the hypothesis that no signal is
present over the background and $H_{sb}$ the hypothesis that signal is
also present. In order to quantify the degree in which each hypotheses
are favoured or excluded by the experimental observation one chooses a
test-statistics which ranks the possible experimental outcomes. A
commonly used test statistics consist as the ratio of the likelihood
function in both hypotheses: $Q=L_{sb}/L_b$ and the quantity $-2\ln Q$
may also be used instead of $Q$. RooStatsCms also provides alternative
choices for the test statistics such as the number of events or the
profiled likelihood ratio.

\begin{figure}[h]
    \begin{center}
        \includegraphics[width=87mm]{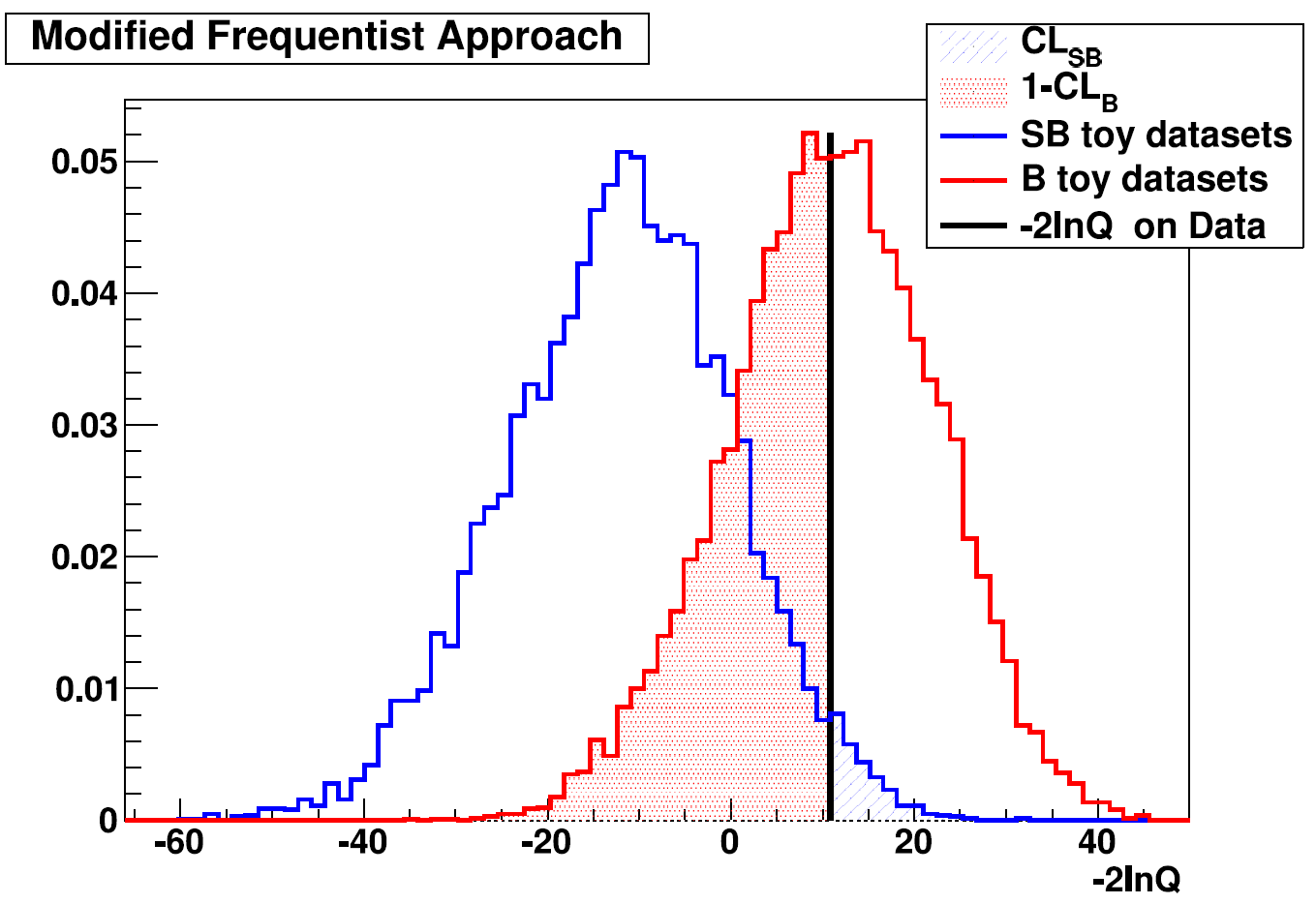}
        \caption{The distributions of $-2\ln Q$ in the background-only
                 (red, on the right) and signal+background (blue,
                 on the left) hypotheses. The black line represents
                 the value of $-2\ln Q$ on the tested data.  The
                 shaded areas represent $1-CL_{b}$ (red) and $CL_{sb}$
                 (blue).\label{m2lnQ}}
    \end{center}
\end{figure}

A comparison of $Q_\mathrm{obs}$ for the data being tested to the
probability distributions $dP/dQ$ expected in both hypotheses allows
to compute the confidence levels:
\begin{eqnarray}
  CL_{sb} = P_{sb}(Q<Q_\mathrm{obs}), & \mathrm{where} & P_{sb}(Q<Q_\mathrm{obs}) = \int_{-\infty}^{Q_\mathrm{obs}}\frac{dP_{sb}}{dQ} dQ, \\
  CL_{b} = P_{b}(Q<Q_\mathrm{obs}), & \mathrm{where} & P_{b}(Q<Q_\mathrm{obs}) = \int_{-\infty}^{Q_\mathrm{obs}}\frac{dP_{b}}{dQ} dQ.
\end{eqnarray}
Small values of $CL_{sb}$ (resp.~$CL_{b}$) point out poor compatibility with the $H_{sb}$ (resp.~$H_{b}$)
hypothesis and favour the $H_{b}$ (resp.~$H_{s}$) hypothesis.
The functional form of the $dP_{sb}/dQ$ and $dP_{b}/dQ$ pdfs not being
known a priori, a large amount of toy Monte-Carlo experiments are performed
in order to determine it for
two families of pseudo datasets: the ones in the signal+background and the ones in 
the background-only hypothesis (see figure~\ref{m2lnQ}).

\begin{figure}[h]
    \begin{center}
        \includegraphics[width=86mm]{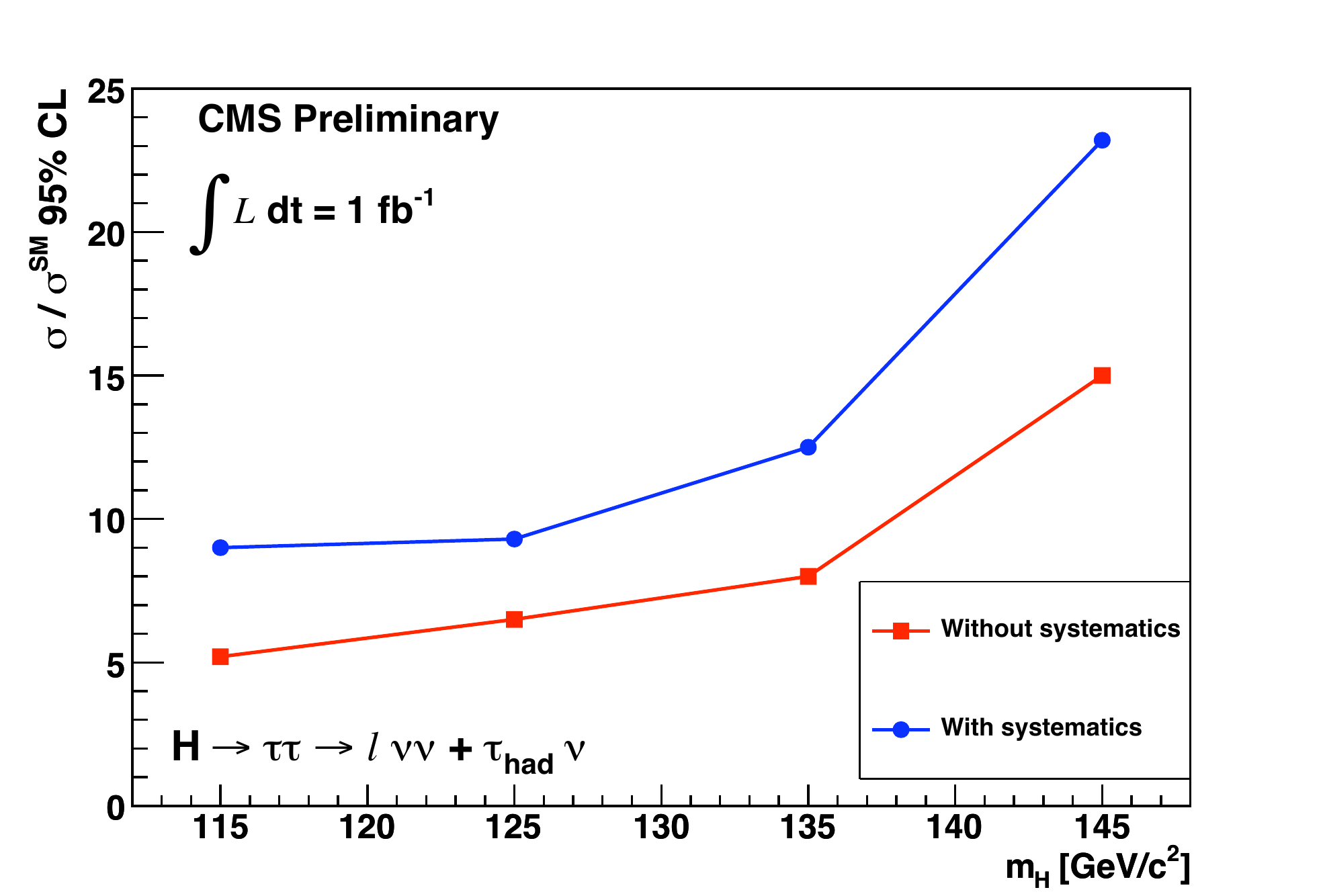}
        \caption{\label{htautau_lim_1}$H\rightarrow\tau\tau$
                 expected exclusion plot \cite{htautau}: limits in presence or 
                 absence of systematics. The effect of the systematics is to 
                 deteriorate the exclusion power of the analysis.}
  \end{center}
\end{figure}

\begin{figure}[h]
    \begin{center}
        \includegraphics[width=86mm]{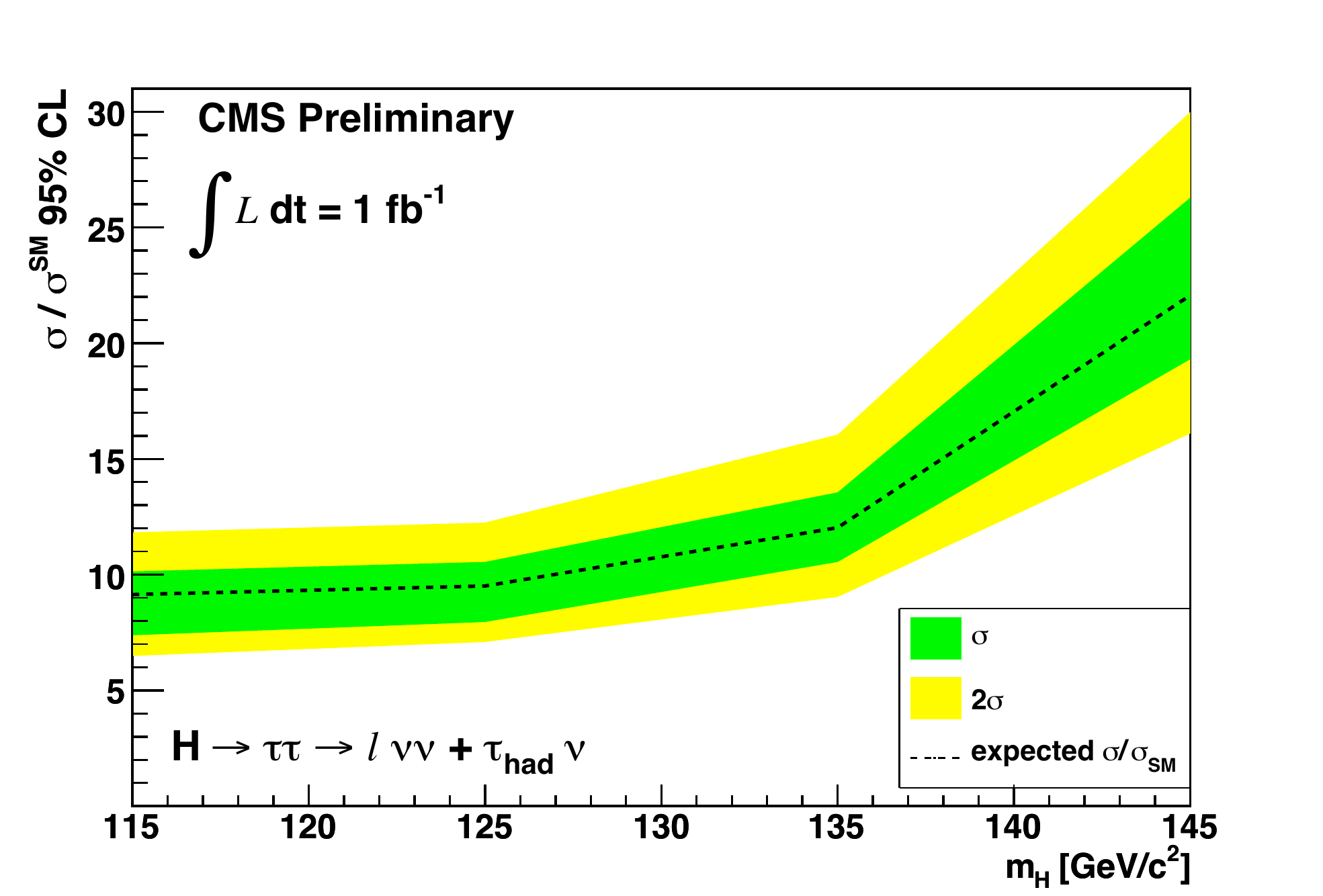}
        \caption{\label{htautau_lim_2}$H\rightarrow\tau\tau$ 
                 expected exclusion plot \cite{htautau}: 
                 The 1$\sigma$ and 2$\sigma$ bands are obtained 
                 assuming a 1$\sigma$ or 2$\sigma$ upwards 
                 (downwards) fluctuation of the number of observed background 
                 events. The plot is produced with the 
\texttt{ExclusionBandPlot} class.}
  \end{center}
\end{figure}

\begin{figure}[h]
    \begin{center}
        \includegraphics[width=86mm]{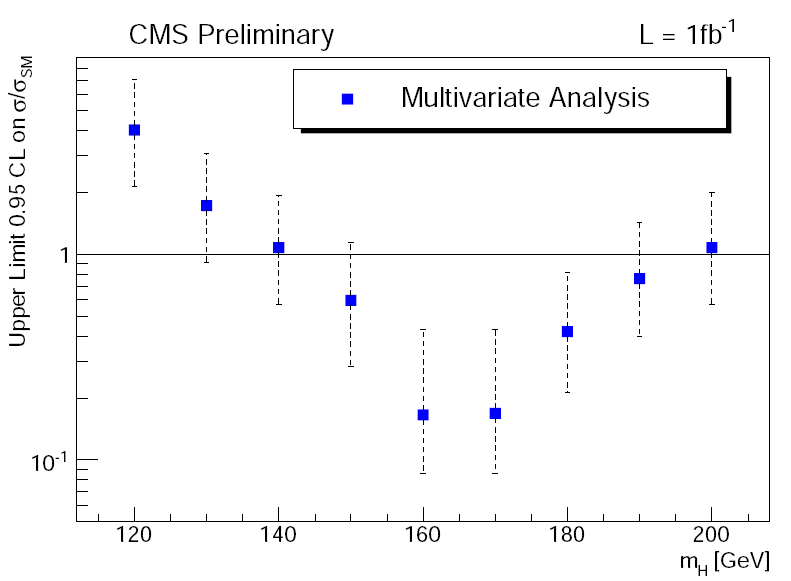}
        \caption{$H\rightarrow WW^{(*)}\rightarrow 2l 2\nu$ \cite{hww}:
                 expected 95\% CL upper limits with the 
                 profile likelihood method for each of the Higgs mass 
                 hypotheses considered (in the assumption of no signal).}\label{hww_lim}
    \end{center}
\end{figure}

\begin{figure}[h]
    \begin{center}
        \includegraphics[width=86mm]{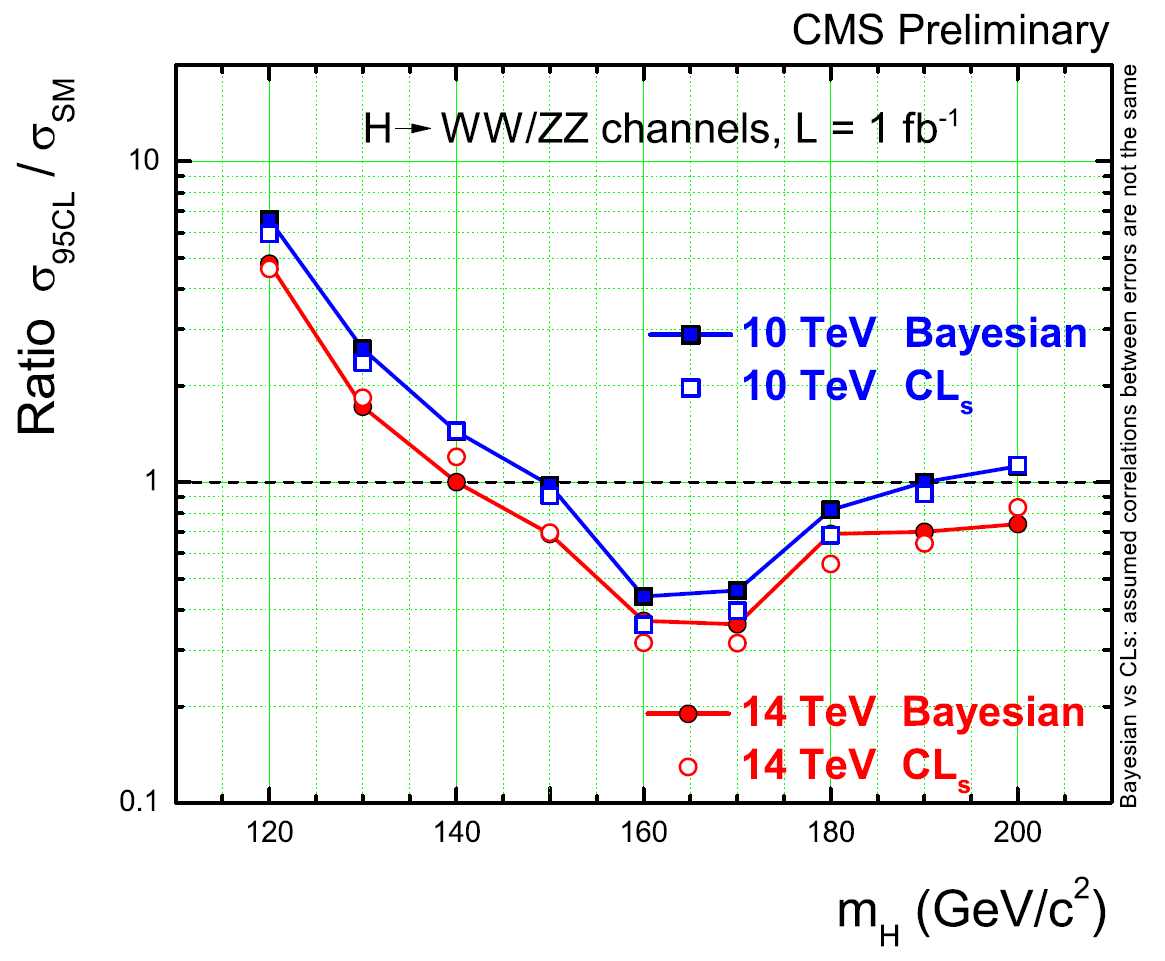}
        \caption{Projected exclusion limits for the Higgs boson at 14 and 
                 10 TeV centre of mass energies: 
                 $H\rightarrow ZZ^{(\ast)}$ and 
                 $H\rightarrow WW^{(\ast)}$ channels \cite{hww,hzz} are 
                 combined with two different methodologies, Bayesian and CL$_s$. 
                 Those results show a very reasonable 
                 agreement (even if the modelling of the constraints and 
                 correlations slightly differ in the two approaches). 
                 The Bayesian combination was carried out with a private tool, 
                 which confirmed the results obtained with RooStatsCms.}\label{combination_plot}
    \end{center}
\end{figure}

A significance estimation can be obtained using formula~\ref{nsigmasfromCL2} on
$CL_{sb}$. Moreover, the tested data can be said to be excluded at a
given CL if $1-CL_{sb}$ is smaller than this CL (or alternatively the
$CL_s$ prescription can be used (see below)). By varying the
hypothesis being tested (for example varying the signal cross-section
as on figures~\ref{htautau_lim_1}, \ref{htautau_lim_2}, \ref{hww_lim}
and \ref{combination_plot}) one may also scan for the type of model
that can be excluded with the given data.
It should be observed that these confidence intervals do not have the
same meaning as the ones obtained with the profile likelihood method
or the Bayesian credibility intervals.

\subsection{The $CL_{s}$ prescription}

Since the hypothesis being tested above is the signal \textit{plus}
background hypothesis and not the signal-only one, it is possible that
unphysical regions are included in the confidence interval obtained, if
the data is affected by large downward fluctuation of the background.
In order to avoid this feature of a pure frequentist approach, the
\textit{modified}, or \textit{conservative}, frequentist approach 
($CL_s$ method) was often used in High Energy Physics (e.g.~by LEP, Tevatron and Hera
experiments). Here, one uses the ratio of p-values, $CL_{s}=CL_{sb}/CL_{b}$, leading
to more conservative limits.
Even if $CL_s$ is not technically a confidence level, the signal
hypothesis is here considered excluded with a certain confidence level
CL when $1-CL_s<\mathrm{CL}$.

\subsection{Inclusion of systematics uncertainties}

Systematics uncertainties can be taken into account by various
techniques. In the likelihood methods described
in section~\ref{Profile_Likelihood} a very convenient approach is to use
the profiling procedure while in the frequentist method of
section~\ref{Frequentist_methods} a Monte-Carlo marginalisation
technique can be applied. Both methods require to assume a
probability distribution for the systematics, or \textit{nuisance}, parameters
(this probability distribution would be called 
the \textit{prior probability} in a Bayesian context). 
RooStatsCms allows, for example, to assume a parameter $\theta_s$ 
to be distributed in an interval
$[\theta_{s,\mathrm{min}};\theta_{s,\mathrm{max}}]$ according to a
Gaussian, a log-normal or a flat distribution.

In the same treatment, it is possible to take into account
correlations between parameters by providing the full covariance
matrix to RSC. While for uncorrelated nuisance parameters, the
global prior probability distribution simply consist of the product of
the individual distributions, when a correlation between $n\geq2$
nuisance parameters is known, one can/should use a $n$-dimensional
prior probability distribution.
The profiling (see section~\ref{Profile_Likelihood}) takes place through the
minimisation of the negative log-likelihood function that could take into 
account the systematics and correlations. 
This does not require any Monte-Carlo integration. 
Suppose that the $\theta_{s}$ parameter is affected by the systematic uncertainties 
described by the $g({\theta_{s}})$ pdf. One writes the joint pdf describing
the data and parameters as 
\begin{equation}
    f'(\underline{x},\underline{\theta})=f(\underline{x},\underline{\theta})\cdot g({\theta_{s}}).
\label{penalised_likelihood}
\end{equation}
In the negative log-likelihood function $g(\theta_s)$ contributes as
an additive penalty term: there is freedom in varying $\theta_s$ but
$-\ln g(\theta_s)$ become large if going too far from the expected
value (w.r.t. the magnitude of the assumed uncertainty).
The scan of this altered likelihood preserves the position of the minimum
point but has in general a larger curvature leading to broader
confidence intervals and less aggressive limits.
Once a dataset is specified, RooStatsCms, in presence of systematics, 
automatically creates a likelihood of the type described in equation
\ref{penalised_likelihood}.

The second approach applies Bayesian Monte-Carlo sampling described 
section~\ref{Frequentist_methods}~\cite{CousinsHighland}. 
It consists in varying for each toy Monte-Carlo experiment the effective value 
of the nuisance parameters before generating the toy Monte-Carlo sample itself 
(that includes in addition the Poisson fluctuations).
The whole phase space of the nuisance parameters is thus sampled through
Monte-Carlo integration. The final net effect consist in a broadening
of the $-\ln Q$ distribution and thus, as expected in presence of
systematic uncertainties, a degraded separation of the hypotheses.

\section{Examples of applications}
\label{Examples_of_applications}

RooStatsCms has been used in different contexts up
to now and also within the CMS collaboration, exemplified here by some of the
public results of Standard Model Higgs boson analyses~\cite{htautau, hww, hzz}.  
Three examples of RSC applications are shown,
two of which comprise a combination of analyses.

\section{Conclusion and outlook}
\label{Conclusions_and_outlook}

RooStatsCms is a tool for analysis modelling, statistical studies and
the combination of analyses. It is based on ROOT and exploits
the RooFit technology. The datacard approach described in section \ref{Analyses_modelling}, provides a common base to describe the
analyses and share the results among the groups.
A number of popular statistical methods are provided
by the package: frequentist and modified frequentist,
profile likelihood, an interface for the BAT package providing
Bayesian interpretation of the data \cite{BAT} and a prototype for an
implementation of the Feldman-Cousins method \cite{FeldmanCousins}.
To perform the very lengthy calculations implied by
some methods, the splitting of processes into sub-jobs and the
recollection of results is eased due to the class
structure. As shown in section
\ref{Examples_of_applications}, CMS has been using RSC
in a number analyses. The implemented methods
have been carefully validated by the analysis groups and analysis
review committees, and the results are publicly available.

RooStatsCms can be also seen, in a wider context, as the starting point of the
CMS contribution to the RooStats~\cite{RooStats} project: a
joint effort of the LHC collaborations and the ROOT team,
oversighted by a committee formed by the ATLAS~\cite{ATLAS} and CMS
statistics committees. RooStats is part of ROOT since the 5.22 release of
December 2008 and is currently in rapid evolution~\cite{MonetaCHEP}.
Parts of RSC are integrated into this first RooStats release, 
and further migration work is on-going.
RSC presently implements more features than this initial
RooStats version, but will be modified to use the future released
versions of RooStats. 
The goal is to provide a common tool for
statistical analysis and the combination of measurements implementing
the methods recommended by the statistics experts of the LHC
collaborations. Furthermore, the tool will allow to store a full
analysis model and transfer it from one working group to another or
between experiments. This will ease enormously the cumbersome task of
combination of experimental results. A full statistical re-analysis of
the combined results, based on the original modelling of the
contributing working groups, becomes possible while taking properly
into account correlated parameters as well as common experimental or
theoretical uncertainties.  The anticipated broad basis of potential
users will improve the reliability and robustness of the tool.  Much
experience towards a common statistics tool for High Energy Physics
has been acquired throughout the process of the development of the
CMS-specific tool RSC and the intense consultation with the
experimental groups was extremely useful for the definition of the
most important features and their implementation, thus set the basis
for a decisive contribution of CMS to ROOT within the RooStats
project. The CMS-specific tool will be adapted to rely on and interface the newly
developed common classes, and will continue to provide a common
interface for analysis modelling to the CMS collaboration.  It will
also be useful as a testing ground for new ideas to be explored within
CMS.

\section*{Acknowledgements}
We wish to thank the Higgs working group of CMS for their readiness to
adopt the newly developed tool and for their valuable feedback. We are
indebted to the statistics committee, and in particular to Bob Cousins
for the exchange of ideas, advice and encouragement. We also wish to
thank Wolfgang Wagner the possibility to validate some of the
implemented methods by comparison to code used in one of the single
top analyses by the CDF experiment. We thank the German Ministry of
Science and Education, BMBF, for partly funding this work.

\end{document}